\documentclass[final]{IEEEtran}

\usepackage{amsthm,amssymb,graphicx,multirow,amsmath,color,amsfonts}
\usepackage[update,prepend]{epstopdf}
\usepackage[noadjust]{cite}
\usepackage[latin1]{inputenc}
\usepackage{tikz}
\usetikzlibrary{arrows,calc}		
\usepackage{bbm} 
\usepackage{pdfpages}
\usepackage{tabulary}
\usepackage{multirow}
\usepackage{comment}
\usepackage[ruled,vlined]{algorithm2e}
\usepackage{etoolbox}
\usepackage{hyperref}

\usepackage[font=small,labelfont=bf]{caption}
\usepackage{subcaption}

\makeatletter
\renewcommand{\SetKwInOut}[2]{%
  \sbox\algocf@inoutbox{\KwSty{#2}\algocf@typo:}%
  \expandafter\ifx\csname InOutSizeDefined\endcsname\relax
    \newcommand\InOutSizeDefined{}\setlength{\inoutsize}{\wd\algocf@inoutbox}%
    \sbox\algocf@inoutbox{\parbox[t]{\inoutsize}{\KwSty{#2}\algocf@typo:\hfill}~}\setlength{\inoutindent}{\wd\algocf@inoutbox}%
  \else
    \ifdim\wd\algocf@inoutbox>\inoutsize%
    \setlength{\inoutsize}{\wd\algocf@inoutbox}%
    \sbox\algocf@inoutbox{\parbox[t]{\inoutsize}{\KwSty{#2}\algocf@typo:\hfill}~}\setlength{\inoutindent}{\wd\algocf@inoutbox}%
    \fi%
  \fi
  \algocf@newcommand{#1}[1]{%
    \ifthenelse{\boolean{algocf@inoutnumbered}}{\relax}{\everypar={\relax}}%
    {\let\\\algocf@newinout\hangindent=\inoutindent\hangafter=1\parbox[t]{\inoutsize}{\KwSty{#2}\algocf@typo:\hfill}~##1\par}%
    \algocf@linesnumbered
  }}%
\makeatother

\SetKwInOut{KwInput}{Input}
\SetKwInOut{KwOutput}{Output}

\def\nb0{{\mathbf{0}}}
\def\nb1{{\mathbf{1}}}

\newtheorem{lemma}{Lemma}

\newtheorem{assumption}{Assumption}
	
\allowdisplaybreaks 
\IEEEoverridecommandlockouts
\begin{document}
\graphicspath{{./Figures/}}
\title{
Intelligent Surface Optimization in Terahertz under Two Manifestations of Molecular Re-radiation
}
\author{ Anish Pradhan,
J. Kartheek Devineni, Harpreet S. Dhillon and Andreas F. Molisch
\thanks{A. Pradhan, J. K. Devineni and H. S. Dhillon  are with Wireless@VT, Department of ECE, Virginia Tech, Blacksburg, VA, USA (email: \{pradhananish1, kartheekdj, hdhillon\}@vt.edu). A. F. Molisch is with the
Wireless Devices and Systems Group, Ming Hsieh Department of Electrical
and Computer Engineering, University of Southern California, Los Angeles,
CA, USA (email: molisch@usc.edu).  This work was supported by U.S. National Science Foundation under Grant ECCS-2030215.
 } 
}

\maketitle

\begin{abstract}
The operation of Terahertz (THz) communication can be significantly impacted by the interaction between the transmitted wave and the molecules in the atmosphere. In particular, it has been observed experimentally that the signal undergoes not only molecular absorption, but also molecular re-radiation. Two extreme modeling assumptions are prevalent in the literature, where the re-radiated energy is modeled in the first as additive Gaussian noise and in the second as a scattered component strongly correlated to the actual signal. Since the exact characterization is still an open problem, we provide in this paper the first comparative study of the performance of a reconfigurable intelligent surface (RIS) assisted THz system under these two extreme models of re-radiation. In particular, we employ an RIS to overcome the large pathloss by creating a virtual line-of-sight (LOS) path. We then develop an optimization framework for this setup and utilize the block-coordinate descent (BCD) method to iteratively optimize both RIS configuration vector and receive beamforming weight resulting in significant throughput gains for the user of interest compared to random RIS configurations. Our results reveal that a slightly better throughput is achieved under the scattering assumption for the molecular re-radiation than the noise assumption.
\end{abstract}


\section{Introduction} \label{sec:intro}

The high data rate requirement of beyond-5G (B5G) and 6G networks has inspired the wireless community to consider the terahertz (THz) frequency range (especially, 0.1-10 THz), where the availability of enormous bandwidth can support the new emerging  applications, such as augmented/virtual reality (AR/VR) \cite{tripathi2021millimeter, tataria20216g}. However, communication in THz channels comes with the following set of challenges which restrict the effective utilization of the available bandwidth: a) high free space path-loss, b) high molecular absorption, and c) low efficiency of diffraction processes, and thus deep shadow fading. 
One of the possible solutions to overcome these challenges is to use reconfigurable intelligent surfaces (RISs). In particular, these low-cost meta-surfaces can be opportunistically deployed to create virtual line-of-sight (LOS) links where the direct LOS links are blocked. However, integrating RIS with THz comes with its own sets of challenges namely: 1) capturing the effect of molecular re-radiation accurately, and 2) joint optimization of RIS and beamforming weights of end nodes for the THz setup. Since the implications of the first challenge are much less understood, which is also the main motivation behind this work, we first discuss it in detail below. 

The process of atomic and molecular absorption and re-radiation has been investigated for many decades in the physics literature under the name of {\em radiation trapping} \cite{molisch1998radiation}. The atoms or molecules are excited to a higher quantum state by absorption of radiation matching the resonance frequency, and then re-radiate energy at the same (or similar) frequency when decaying to the ground state. Several frequency bands at which absorption/reemission occurs exist in the THz spectrum, associated with different atmospheric molecules \cite{discussion}. This re-radiation is generally modeled as additive noise in the existing THz literature  by linking it with \textit{sky-noise} \cite{jornet_2012}. To the best of our knowledge, this assumption has not been properly validated by any measurement studies. In addition, there is a second school of thought, which models this phenomenon as scattering, where the existence of multiple scattered copies of the signal due to re-radiation has some basis in the literature \cite{Harde1, Harde2}. Following that, \cite{molabs} recently modeled the THz channel as a Rician channel, where the Rician factor is calculated from the molecular absorption coefficient. Please note that this scattering could also lead to delay dispersion \cite{discussion}, but that is beyond the scope of this paper. Both manifestations have been used in the literature, and it is not straightforward to determine the extent to which each effect is prevalent. To be precise, the true effect will probably lie in between as perhaps a superposition of these two extreme cases. In the absence of extensive measurements, there is no way to characterize this true effect. Given this, the best one can do is to investigate the two extreme scenarios, and characterize their impact on the performance of RIS.

Although the relevant literature on RIS \cite{wu2019beamforming,RISCR,ye2020joint} in the conventional microwave spectrum deal with the joint optimization of RIS configuration vector and beamforming weights, they are not exactly applicable in this case due to the peculiar characteristics of the THz channels. On the other hand, the RIS literature that considers the THz band, such as \cite{VRTHz1,VRTHz2,VRTHz3}, either ignore the different manifestations of the molecular re-radiation, or the dependence of the receiver noise on this re-radiation and the phase-shift configuration of RIS. We fill this gap by considering a parametric model for the THz channel that handles the two extreme effects of re-radiation, and formulate a joint RIS-receive beamformer optimization framework for the proposed channel model. To the best of our knowledge, this is the first paper that compares a jointly optimized multi-antenna system in a THz environment with two extreme assumptions about molecular re-radiation.


\subsection{Contributions}
Assuming an ad-hoc setup in the THz band, we study the performance of a single-antenna transmitter (Tx) associated with an RIS-assisted multi-antenna receiver (Rx) in the presence of other single-antenna interferers. Our main contributions in this work can be summarized as: 1) a novel parametric THz channel that takes into consideration the two extreme manifestations of the re-radiation in the THz band, 2) the characterization of the dependence of molecular re-radiation noise on the RIS configuration, 3) the joint optimization of RIS, and receive beam-forming vector to maximize throughput using the BCD algorithm, and 4) comparing the performance of the optimized system under the two modeling assumptions regarding the nature of molecular re-radiation. Comparing their performance with optimized RIS and receiver beamformer will inform us about when we should care about the exact nature of the re-radiation. As expected, our results  reveal that when the re-radiation is assumed as scattering, the throughput of the corresponding optimized system is slightly higher compared to when it is assumed as noise.

\section{System Model} \label{sec:SysMod}

\begin{figure}
    \centering
    \includegraphics[width=0.9
    \linewidth]{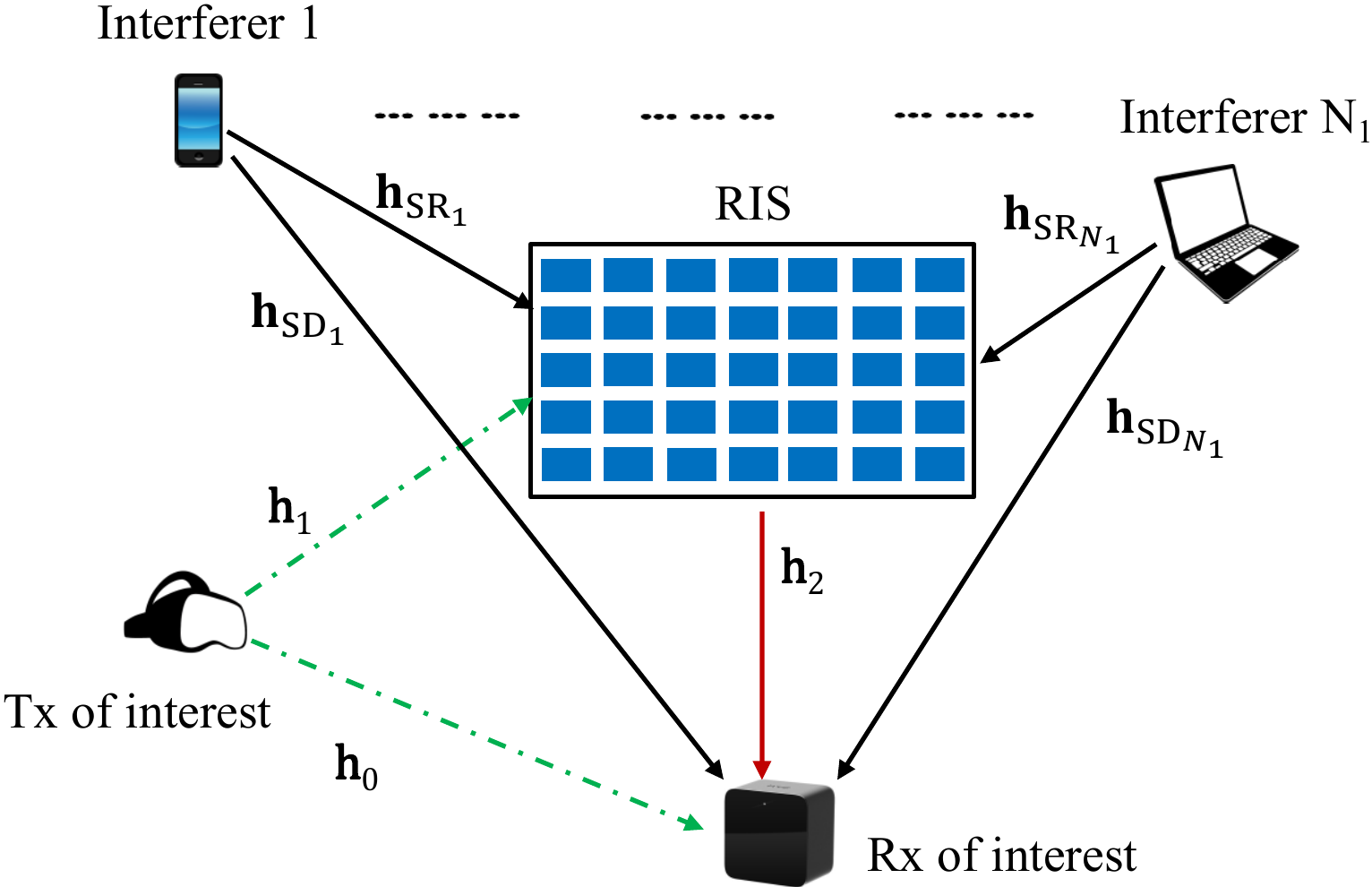}
    \caption{System model for the RIS-assisted communication in THz.}
    \label{fig:sysmodel2}
\end{figure}

We consider an ad-hoc network setup with multiple Tx-Rx pairs communicating simultaneously in the same frequency band. The Tx of interest 
(${\rm Tx}_0$) is assumed to be a single-antenna device, while the Rx of interest (${\rm Rx}_0$) is considered to have multiple antennas and is assisted by a reconfigurable passive surface. The multi-antenna ${\rm Rx}_0$ has $N_R$ receive antennas, while the RIS has $N$ elements. Additionally, there are $N_I$ co-channel single-antenna users that interfere with ${\rm Tx}_0$. The number of elements $N$ at RIS is assumed to be much larger than $N_R$ and $N_I$. It is assumed that each of the communicating devices can have two paths to ${\rm Rx}_0$, one link coming directly from the Tx, and another link reflected from the RIS. The described system model is illustrated in Fig. \ref{fig:sysmodel2}. Note that the $N_I$ interfering users are communicating with their own Rxs, which do not affect our analysis and are therefore not included in Fig. \ref{fig:sysmodel2}. Both RIS and ${\rm Rx}_0$ are assumed to have uniform linear arrays (ULAs) with half wavelength spacing. The array factor of a general ULA is defined as follows:
\begin{align}
 \mathbf{a}_{N_0}({\theta})&=\begin{bmatrix}
1 & e^{j2\pi\frac{d}{\lambda}\sin{{\theta}}} & \hdots & e^{j2\pi\frac{d}{\lambda}(N_0-1)\sin{{\theta}}}
\end{bmatrix},
\end{align}
where $\frac{d}{\lambda}$ is the wavelength spacing, ${N_0}$ is the number of elements of that ULA, and $\theta$ is the angle of arrival or departure depending on the context.

\subsection{Terahertz Channel Model}

\textcolor{black}{Due to the molecular absorption phenomenon in the THz band, a fraction of the propagated signal gets absorbed. The fractions of the signal power for the component available to the receiver node and the component absorbed are respectively given by $\tau(f,d)$ and $1-\tau(f,d)$, where $\tau(f,d)=e^{-k(f)d}$ is the transmittance of the channel, $k(f)$ is the molecular absorption coefficient, $f$ is the operating frequency, and $d$ is the link distance. The absorption coefficient is calculated using the equation $k(f)=\sum_i y_i(f,\mu)+g(f,\mu)$ \cite{SimpleTHz}, where $\mu$ is the mixing ratio of the water vapor present, $y_i$ is the absorption coefficient for the $i$-th absorption line, and $g$ is an error-correcting polynomial. Details on these functions can be readily found in \cite{SimpleTHz} and are hence skipped here due to space limitations. Recall from Section~\ref{sec:intro} that the absorbed power is radiated back in the same band, and two extreme modeling assumptions are prevalent about the manifestation of this molecular re-radiation. This re-radiation happens almost isotropically and the energy is distributed in all the directions for the THz band (unlike the forward scattering we see in visible light due to small particles) \cite{vanExter:89}. In the literature, however, it is commonly assumed that the whole absorbed power is available at the Rx node through re-radiation \cite{jornet_2012,molabs}. Even though this is never explicitly mentioned, this follows from the assumptions that a single absorption/re-radiation event happens over the whole propagation path, and that all of the re-radiated power is scattered in the direction of the Rx node. We use these assumptions in this work as well. For a systematic exposition, we now mention the assumptions related to the two extremes of molecular re-radiation next.}
\begin{assumption} \label{assumption:noise}
\textcolor{black}{Molecular re-radiation is modeled as additive Gaussian noise.} 
\end{assumption}

\begin{assumption} \label{assumption:scatter}
\textcolor{black}{Molecular re-radiation is modeled as a scattering event where the channel response includes an NLOS component.} 
\end{assumption}
\textcolor{black}{
In both of these assumptions, the fraction of power of the additive noise and the NLOS component is equal to $1-\tau(f,d)$ that of the total signal power.
Due to the nature of their modeling, the two assumptions lie on the two ends of the spectrum with the reality lying somewhere in between.
For Assumption \ref{assumption:scatter}, the Rician factor $K_d$, for link distance $d$, is
\begin{align}
    K_d=\frac{\text{Power of the LOS channel}}{\text{Power of the NLOS channel}}=\frac{\tau(f,d)}{1-\tau(f,d)}.
\end{align}}
For analytical convenience, we unify both Assumptions \ref{assumption:noise} and \ref{assumption:scatter} about the re-radiation phenomenon by incorporating a variable $\zeta$ in the conventional Rician channel model \cite{molabs}. We note that, in the first case, the unified model should consist of just the LOS component as the re-radiation signal manifests as Gaussian noise, whereas in the second case, the channel model should consist of both the LOS and NLOS components due to the manifestation of re-radiation as scattering. This is captured in the model by configuring $\zeta=1$ and $\zeta=0$ for Assumption \ref{assumption:noise} and \ref{assumption:scatter}, respectively. Further details on the modified Rician channel with the $\zeta$ parameter are provided next.

We denote the channel responses between $\mathrm{X}$ and $\mathrm{Y}$ by this unified formulation below:
\begin{align}
    &\mathbf{h}_{\mathrm{XY}}\!=\!\left(\sqrt{\frac{K_{\bar{d}}}{K_{\bar{d}}+1}}\mathbf{F}_{\mathrm{LOS}}+\sqrt{\frac{1-\zeta}{K_{\bar{d}}+1}}\Tilde{\mathbf{h}}_{{XY}}\right)\frac{c}{4\pi f \bar{d}}, \label{eq:directchannel}
\end{align}
where $\mathrm{X=S\text{ or }R}$ denotes RIS or the Rx, $\mathrm{Y=R\text{ or }}\mathrm{T}_i$ denotes the Rx or the $i$-th Tx, $\bar{d}$ is the link distance between $\mathrm{X}$ and $\mathrm{Y}$, $\Tilde{\mathbf{h}}_{{XY}}$ is the NLOS counterpart and is a complex matrix of the same dimension as $\mathbf{h}_{\mathrm{XY}}$, where each element is an independent and identically distributed (i.i.d.) complex normal random variable with zero mean and unit variance, and $f$ is the operating frequency. \textcolor{black}{Note that, when $\zeta=1$, the NLOS component vanishes and $\frac{K_{\bar{d}}}{K_{\bar{d}}+1}=\tau(f,d)$ corresponds to the transmittance as expected.} The different channel links in the system are defined through the next table:
\renewcommand{\arraystretch}{1.3}
\begin{center}
\begin{tabular}{ |c||c|c|c| } 
 \hline
 $\mathrm{XY}$ & $\mathrm{RT}_i$ & $\mathrm{ST}_i$ & $\mathrm{SR}$ \\ 
 \hline
 $\bar{d}$ & $d_i$ & $d_{\gamma_i}$ & $d_\alpha$ \\ 
 \hline
 $\mathbf{F}_\mathrm{LOS}$ & $\mathbf{a}_{N_R}^H({\theta}_{R,i})$ & $\mathbf{a}_{N}^H({\theta}_{S,i})$ & $\mathbf{a}_{N_R}^H({\theta}_\alpha)\mathbf{a}_{N}({\theta}_\beta)$ \\ 
 \hline
\end{tabular}
\end{center}
For notational ease, we define the following stacked channels:
\begin{align}
&\mathbf{H}_{i}=[\mathbf{h}_\mathrm{SR}\mathrm{Diag}\mathbf{h}_{\mathrm{ST}_{i}}\hspace{5pt}\mathbf{h}_{\mathrm{RT}_{i}}]~ \forall i\in [0,N_I].  \label{eq:stacked}
\end{align}

\subsection{Signal Model}
Assuming that the $i$-th Tx transmits a signal $x_{i}$ with power $\mathrm{E}[|x_i|^2]=P_i$, the received signal at ${\rm Rx}_0$ can be given as:
\begin{align}
    \mathbf{y}=&(\mathbf{h}_{\mathrm{RT}_0}+\mathbf{h}_\mathrm{SR}\mathrm{Diag}\mathbf{h}_{\mathrm{ST}_0}\mathbf{\Theta}^T)x_0+ \notag
    \\ &
    \sum\limits_{i=1}^{N_I}(\mathbf{h}_{\mathrm{RT}_{i}}+\mathbf{h}_\mathrm{SR}\mathrm{Diag}\mathbf{h}_{\mathrm{ST}_{i}}\mathbf{\Theta}^T){x}_i + \mathbf{w}, \label{eq:signalmodel0}
\end{align}
where $ \mathbf{\Theta} = [
\varphi_1\,
\hdots \,
\varphi_N]$ is the RIS configuration vector, $\varphi_n \in \mathbb{C}$ is the $n$-th RIS element's reflection coefficient for all $n\in\{1,\hdots,N\}$ with $|\varphi_n|\leq1$, and $\mathbf{w}$ is the additive noise signal with variance $\sigma_w^2 + \zeta\sigma_m^2$. The variance parameters $\sigma_w^2$ and $\zeta\sigma_m^2$ correspond to the thermal noise and molecular re-radiation noise, respectively. Under Assumption \ref{assumption:noise}, the molecular re-radiation noise variance is $\sigma_m^2=\sum\limits_{i=0}^{N_I}\sigma_{m,i}^2$, where $\sigma_{m,i}^2$ is the sum of the individual molecular re-radiation noise due to ${\rm Tx}_i$. Note that when $\zeta=0$, the noise variance of the molecular re-radiation vanishes and manifests in fading since it corresponds to Assumption \ref{assumption:scatter}. 

Defining $\mathbf{\Theta}_0=[\mathbf{\Theta}\hspace{5pt}1]^T$, and using the stacked channel definition from (\ref{eq:stacked}), we can rewrite the received signal as:
\begin{align}
&\mathbf{y}=\mathbf{H}_0\mathbf{\Theta}_0x_0+\sum\limits_{i=1}^{N_I}\mathbf{H}_{i}\mathbf{\Theta}_0x_i + \mathbf{w}.\label{eq:signal}
\end{align}
Multiplying the received signal with the receive beamformer $\mathbf{u}$, the corresponding signal to noise ratio (SINR) $\gamma$ is:
\begin{align}
&\gamma(\mathbf{u},\mathbf{\Theta}_0)=\frac{P_0|\mathbf{u}^H\mathbf{H}_0\mathbf{\Theta}_0|^2}{\sum\limits_{i=1}^{N_I}P_i|\mathbf{u}^H \mathbf{H}_{i}\mathbf{\Theta}_0|^2+\mathbf{u}^H\mathbf{u}(\sigma_w^2+\zeta\sigma_{m}^2)},\label{eq:SINRexpression}
\end{align}
\begin{lemma} \label{lem:NoiseRISdependence}
\textcolor{black}{The variance of the molecular re-radiation signal component modeled as additive Gaussian noise is given by  $\zeta\sigma_{m,i}^2$ for the ${\rm Tx}_i$, where $\sigma_{m,i}^2=\sigma_{m_1,i}^2+\sigma_{m_2,i}^2\mathbf{\Theta}_0^H\mathbf{\Theta}_0$, $\sigma_{m_1,i}^2=\left(\frac{c}{4\pi f d_i}\right)^2P_i[1-\tau(f,d_i)]$, $\sigma_{m_2,i}^2=\left(\frac{c^2}{16(\pi f)^2 }\frac{1}{d_\alpha d_{\gamma_i}}\right)^2P_i[1-\tau(f,d_\alpha)\tau(f,d_{\gamma_i})]$, $P_i$ is the transmit power, $d_\alpha$ is the distance between ${\rm Rx}_0$ and RIS, and $d_{\gamma_i}$ is the distance between RIS and ${\rm Tx}_i$.
\begin{IEEEproof}
See Appendix \ref{sec:Lem1Proof}.
\end{IEEEproof}}
\end{lemma}
From the Lemma \ref{lem:NoiseRISdependence}, we define $\sigma_{m_1}^2=\sum\limits_{i=0}^{N_I}\sigma_{m_1,i}^2$, and $\sigma_{m_2}^2=\sum\limits_{i=0}^{N_I}\sigma_{m_2,i}^2$ for notational convenience.
\section{Joint Optimization of Receive Beamformer and RIS Configuration Vector} \label{sec:OpProblems}
In this section, the joint optimization of the receive beamforming weight and the RIS configuration vector is studied. We assume that the perfect channel state information (CSI) is available at the receiver. As the maximization of  \eqref{eq:SINRexpression} is a non-convex optimization problem over the two coupled variables, we use the BCD algorithm to solve the joint optimization problem by splitting it into two sub-problems, one for each variable, and solving each of the sub-problem in alternating steps which results in an efficient solution. The first sub-problem can be conveniently formulated as a maximization of the generalized Rayleigh quotient, and thus has a straightforward analytical solution. The second sub-problem is approximately solved by formulating it as a rank-relaxed semidefinite programming (SDP) problem. Using the bisection method, we first find a higher rank solution to the SDP problem, and then obtain a rank-one solution through the Gaussian randomization procedure for the original sub-problem. Finally, we prove the convergence of the proposed algorithm.
\subsection{Problem Formulation}
The main objective in our problem is to maximize the achievable throughput of ${\rm Tx}_0$, while making sure that the RIS is passive i.e., $|\varphi_n|<1 , \forall n$. The transmit power $P_0$ is not considered as an optimization variable since the optimal transmit power that will maximize the objective is simply the maximum allowable power, because of which it is simply set to the maximum allowable value. The optimization problem for the throughput maximization can be formulated as:
\begin{subequations}
\label{eq:sinropt}
\begin{align}
\max_{\mathbf{u},\mathbf{\Theta}_0} \quad & \log(1+\gamma\left(\mathbf{u},\mathbf{\Theta}_0\right)) \label{eq:sinroptobj}\\
\textrm{s.t.} \quad & |[\mathbf{\Theta}_0]_l| \leq 1, \forall \, l \in 1,2,\hdots,N+1. \label{eq:sinroptcons}
\end{align}
\end{subequations}
Note that the objective is equivalent to maximizing $\gamma\left(\mathbf{u},\mathbf{\Theta}_0\right)$ due to the monotonically increasing nature of the logarithm.
\subsection{Receive Beamformer Optimization}
In the first sub-problem, the SINR maximization conditioned on a given $\mathbf{\Theta}_0$ can be reformulated as an unconstrained maximization of a generalized Rayleigh quotient:
\begin{align}
\max_{\mathbf{u}} \quad & \frac{\frac{P_0}{\sigma_w^2+\zeta\sigma_m^2}\mathbf{u}^H\mathbf{G}_0\mathbf{G}_0^H\mathbf{u}}{\mathbf{u}^H(\sum\limits_{i=1}^{N_I}\frac{P_i}{\sigma_w^2+\zeta\sigma_{m}^2)} \mathbf{G}_{i}\mathbf{G}_{i}^H+\mathbf{I}_{N_R})\mathbf{u}}, \label{eq:RxBfOp}
\end{align}
where $\mathbf{G}_i=\mathbf{H}_{i}\mathbf{\Theta}_0$. The analytical expression of the optimal receive beamforming weight is given by:
\begin{align}
\mathbf{u}^*=\frac{(\sum\limits_{i=1}^{N_I}\frac{P_i}{\sigma_w^2+\zeta\sigma_m^2} \mathbf{G}_{i}\mathbf{G}_{i}^H+\mathbf{I}_{N_R})^{-1} \mathbf{G}_0}{\left\|(\sum\limits_{i=1}^{N_I}\frac{P_i}{\sigma_w^2+\zeta\sigma_m^2} \mathbf{G}_{i}\mathbf{G}_{i}^H+\mathbf{I}_{N_R})^{-1} \mathbf{G}_0\right\|}, \label{eq:RxBfOptimum}
\end{align}
where the expression is normalized to ensure $\mathbf{u}^H\mathbf{u}=1$.
\subsection{RIS Optimization}

In the second sub-problem, the optimization problem is reformulated with a new optimization variable $\mathbf{\Psi}=\mathbf{\Theta}_0\mathbf{\Theta}_0^H$. Redefining $\mathbf{\Psi}$ helps us to structure the optimization problem in a more tractable form but it introduces two more constraints. The first constraint is that $\mathbf{\Psi}$ has to be a positive semidefinite matrix, which is a convex constraint. However, the second constraint is the rank one constraint on $\mathbf{\Psi}$, which is relaxed in the problem formulated below. Conditioned on a given $\mathbf{u}$, the rank-relaxation problem can be formulated as follows:
\begin{align}
\max_{\mathbf{\Psi}} \quad & \frac{\mathrm{Tr}({\mathbf{\Psi}{\mathbf{L}_{0}}})
}{\mathrm{Tr}(\mathbf{\Psi}\mathbf{\mathbf{M}})+\alpha}
\nonumber\\
\textrm{s.t.} \quad & 
\mathbf{\Psi} \succeq 0,
\nonumber\\
\quad & \left[\mathbf{\Psi}\right]_{l,l} \leq 1, \quad \forall l=1,2,\hdots,N+1, \label{eq:RISmain}
\end{align}
where $\mathbf{L}_i=\frac{P_i}{\sigma_w^2}\mathbf{F}_i\mathbf{F}_i^H$, $\mathbf{F}_i=\mathbf{H}_{i}^H\mathbf{u}$, $\mathbf{M}=\sum\limits_{i=1}^{N_I}\mathbf{L}_i+\zeta\frac{\sigma_{m_2}^2}{\sigma_w^2}\mathbf{I}_{N+1}$, and $\alpha=1+\frac{\zeta\sigma_{m_1}^2}{\sigma_w^2}$.

We further change the optimization problem by introducing a new variable $t \geq 0$ to change (\ref{eq:RISmain}) to an epigraph \cite{Palomar_Eldar_2009}:
\begin{align}
\gamma^*=\max_{\mathbf{\Psi},t\geq 0} \quad & t
\nonumber\\
\textrm{s.t.} \quad & \mathrm{Tr}({\mathbf{\Psi}{\mathbf{L}_{0}}}) \geq t\mathrm{Tr}(\mathbf{\Psi}\mathbf{\mathbf{M}})+t\alpha,
\nonumber \\
\quad & \mathbf{\Psi} \succeq 0,
\nonumber\\
\quad & \left[\mathbf{\Psi}\right]_{l,l} \leq 1, \forall l=1,2,\hdots,N+1,
\end{align}

With $t \geq 0$, the feasibility problem is given by:
\begin{align}
\mathrm{Find} \quad & \mathbf{\Psi}
\nonumber\\
\textrm{s.t.} \quad & \mathrm{Tr}({\mathbf{\Psi}{\mathbf{L}_{0}}}) \geq t\mathrm{Tr}(\mathbf{\Psi}\mathbf{\mathbf{M}})+t\alpha,
\nonumber \\
\quad & \mathbf{\Psi} \succeq 0,
\nonumber\\
\quad & \left[\mathbf{\Psi}\right]_{l,l} \leq 1, \forall l=1,2,\hdots,N+1. \label{eq:RISOptProblem}
\end{align}

The feasibility of the above problem implies that $\gamma^*\geq t$. The infeasibility implies $\gamma^*< t$. From this observation, one can easily solve this using a bisection method. The feasibility problem itself is solved by using CVX \cite{cvx,gb08}. Then, we find a feasible rank-one solution $\mathbf{\Theta}_0^*$ from the optimal $\mathbf{\Psi}^*$ using Gaussian randomization \cite{Palomar_Eldar_2009}. We substitute the last element of $\mathbf{\Theta}_0^*$ by unity to ensure its feasibility.

\begin{algorithm}
\SetAlgoLined
\KwInput{$\mathbf{H}_{i},\zeta~\forall i$}
\KwOutput{$\mathbf{\Theta}_0^*,\mathbf{u}^*$} 
Initialize $\mathbf{\Theta}_0$ with an all-one vector, $i=0$, $\gamma_0=0$, and $\Delta=\epsilon+1$.\\
 \While{$\Delta>\epsilon$}{
  Obtain $\mathbf{u}_{i+1}$ from (\ref{eq:RxBfOptimum}).\\
  Obtain $\mathbf{\Psi}_{i+1}$ by using bisection on (\ref{eq:RISOptProblem}).\\
  Generate $G$ feasible solutions for $\mathbf{\Theta}_{0_{i+1}}$ through Gaussian randomization \cite{Palomar_Eldar_2009}.\\
  Choose the solution $\mathbf{\Theta}_{0_{i+1}}$ that provides the highest $\gamma_{i+1}$ through (\ref{eq:SINRexpression}).\\
 \eIf{$\gamma_{i+1}>\gamma_i$}{
   $\mathbf{\Theta}_{0_{i+1}}^*=\mathbf{\Theta}_{0_{i+1}}$.\\
   }{$\mathbf{\Theta}_{0_{i+1}}^*=\mathbf{\Theta}_{0_{i}}$.\\
   }
   {
  
  }
  Evaluate $\Delta=|\gamma_{i+1}-\gamma_{i}|/\gamma_{i}$.\\
  $i=i+1$.
 }
 \caption{Joint Optimization by BCD} \label{algo:Algo1}
\end{algorithm}
The algorithm ensures that the objective function is non-decreasing each iteration and because it is upper bounded by some value, the alternating algorithm will converge.

\begin{figure}
     \centering
     \begin{subfigure}[b]{0.5\textwidth}
         \centering
         \includegraphics[width=\textwidth]{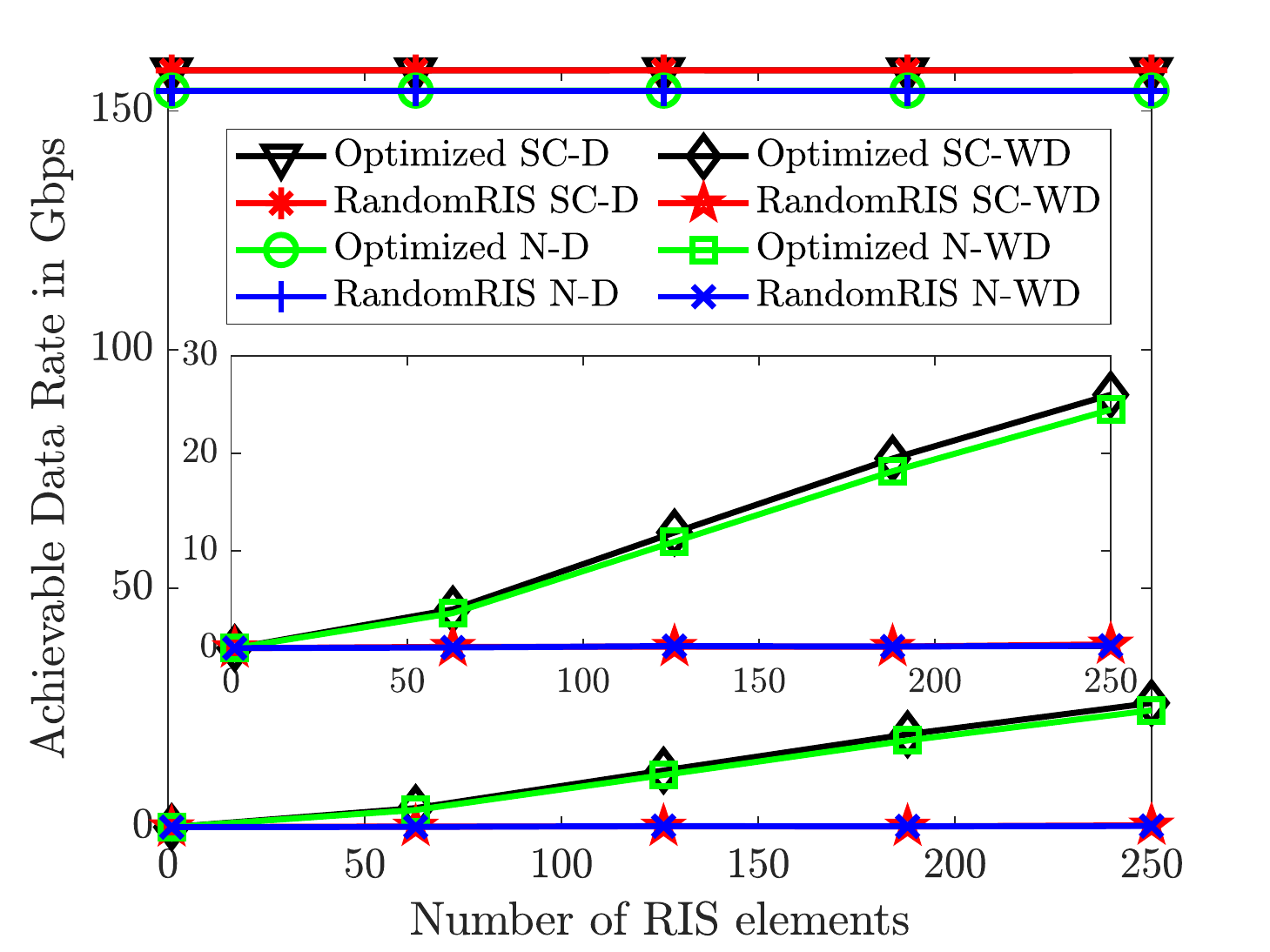}
         \caption{}
         \label{fig:capel}
     \end{subfigure}
     \hfill
     \begin{subfigure}[b]{0.5\textwidth}
         \centering
         \includegraphics[width=\textwidth]{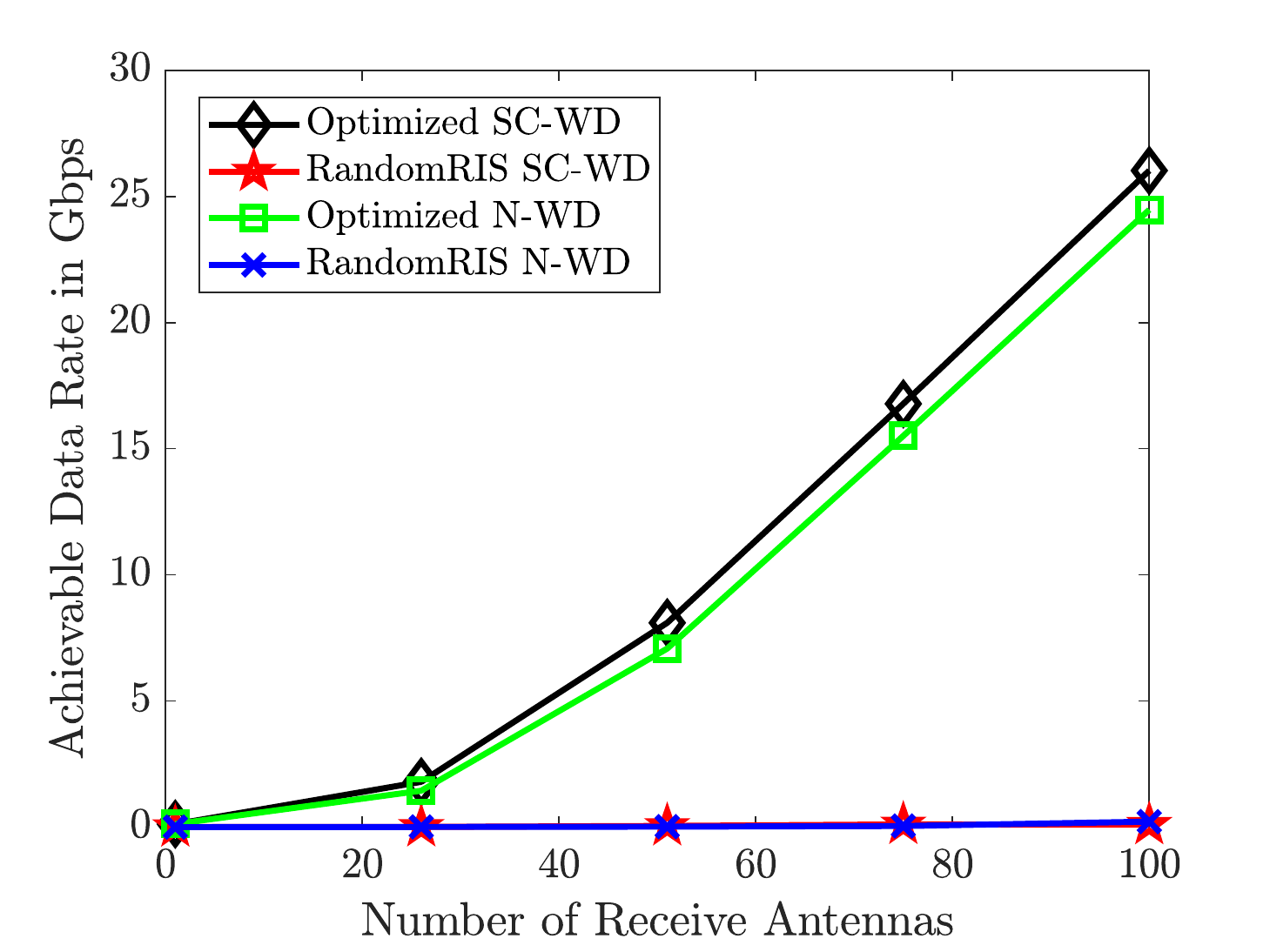}
         \caption{}
         \label{fig:capantwd}
     \end{subfigure}
        \caption{ (a) Achievable throughput with RIS elements, (b) Achievable throughput with Rx antennas without the direct link.}
        \label{fig:capRISant}
\end{figure}

\section{Numerical Results} \label{sec:NumResults}

In this section, we numerically evaluate the performance of our proposed optimization algorithm. The simulation scenario is inspired by a practical indoor VR gaming setup. Three interferers are placed on a circular ring of radius $6$m at $5^o$, $75^o$ and $135^o$ while the ${\rm Rx}_0$ is placed at the center (origin) and is equipped with $100$ antennas. The RIS is situated at a distance of $1\text{m}$ along the positive x-axis and assumed to contain $250$ elements. The ${\rm Tx}_0$ is placed at a distance of $1$m away from the origin at an angle of $60^\circ$ with the positive x-axis and  assumed to transmit at $2$ watt power over the large bandwidth (which is a reasonable assumption, e.g., see \cite{power2w}). All the interferers are also transmitting at the same power. The system parameters, unless otherwise stated, for the simulation setup are as follows: temperature of $27^\circ$C, standard atmospheric pressure of $1$ atm, and relative humidity of $50\%$, the transmission carrier frequency of $220$ GHz, and bandwidth of $10$ GHz. These parameters also dictate the value of $k(f)$. In addition, the thermal noise variance is considered to be $-174$ dBm/Hz. For the BCD algorithm, the parameters $\epsilon, \epsilon_0, \epsilon_1$, and $G$ are set to $10^{-5}, 10^{10}, 10^{-5}$, and $5000$, respectively. 

\begin{figure}
     \centering
     \begin{subfigure}[b]{0.5\textwidth}
         \centering
         \includegraphics[width=\textwidth]{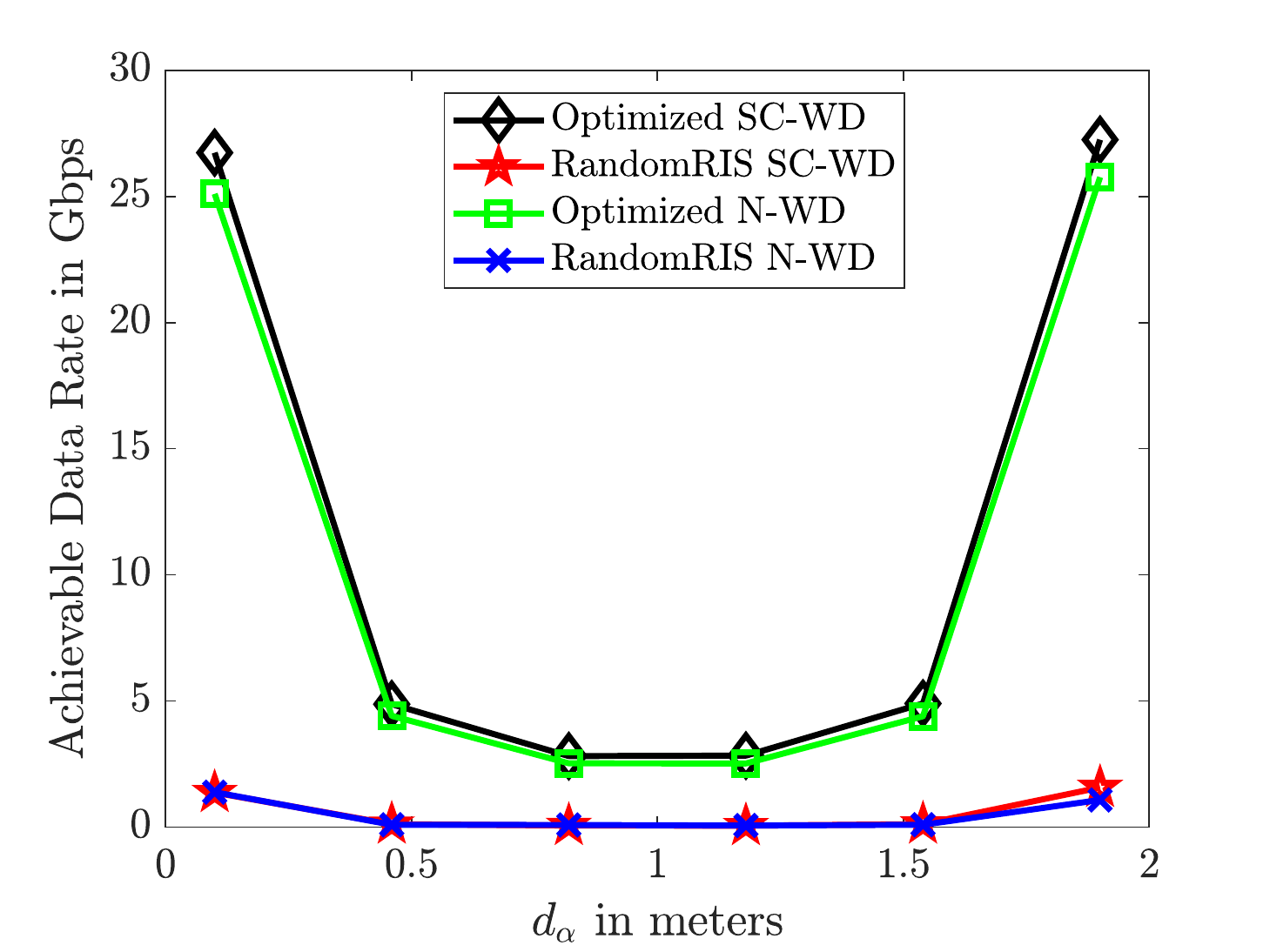}
         \caption{}
         \label{fig:cappos}
     \end{subfigure}
     \hfill
     \begin{subfigure}[b]{0.5\textwidth}
         \centering
         \includegraphics[width=\textwidth]{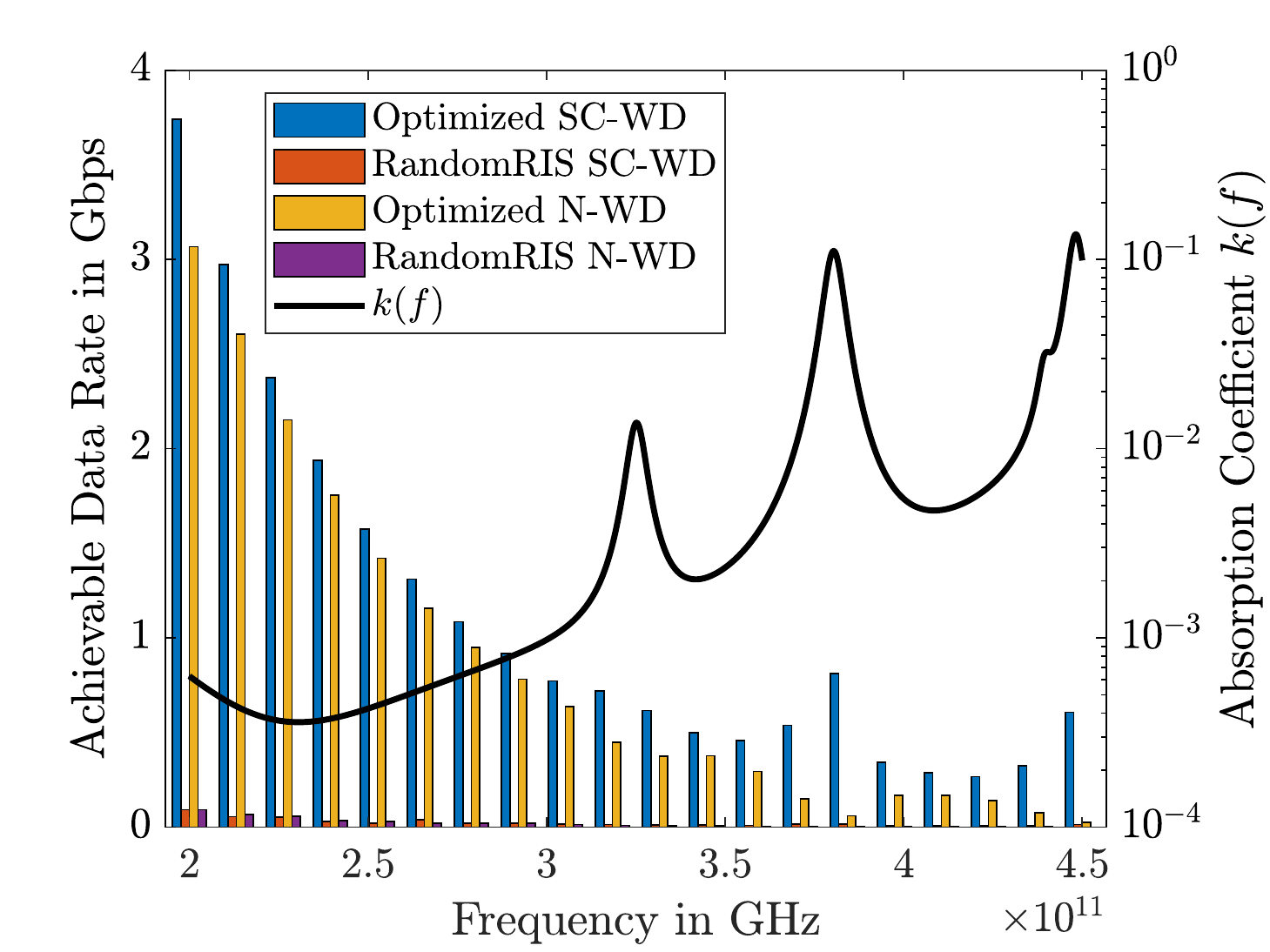}
         \caption{}
         \label{fig:capfwd}
     \end{subfigure}
        \caption{Achievable throughput without the direct link with (a) RIS position, (b) frequency.}
        \label{fig:capdf}
\end{figure}

First, we vary the number of RIS elements and numerically evaluate the achievable throughput. The notations `WD' and `D' denote the two cases depending on the availability of the direct link, and termed as \emph{without direct link} and \emph{with direct link}, respectively. \textcolor{black}{Similarly, the abbreviations `SC' and `N' denote Assumptions \ref{assumption:noise} and \ref{assumption:scatter}, respectively.}  Additionally, the results for both random and optimized configurations of the RIS are shown in the numerical plots. As we can see from Fig. \ref{fig:capel}, when the direct link is available, the RIS does not significantly enhance the performance even with $250$ elements. In the same Fig. \ref{fig:capel}, we show a zoomed-in version focusing on plots for the case without the direct link, which illustrates a significant throughput gain that is achieved with the optimal RIS configuration. This use-case corresponds to indoor communications, and its importance is emphasized by the fact that LOS can be lost due to blockage from the user's own body (self-blockage) or from other user's body (dynamic blockage). Here, the optimal configuration of RIS provides a gain of almost $25$ Gbps in throughput over the random configuration that underscores the impact of RIS. Another important observation from Fig. \ref{fig:capel} is that a slightly lower throughput gain is achieved with the optimization process under Assumption \ref{assumption:noise}. Intuitively, this can be explained by the larger noise variance and loss in signal power. The small gap between these assumptions imply that the results for this particular setup are not sensitive to a specific assumption. As the molecular re-radiation depends on the transmit power, the gap may become more prominent with the increase of transmit power. Similar trend is observed in Fig. \ref{fig:capantwd} where we look at the impact of increase in the number of receive antennas on the throughput gains. Note that the absolute values of the achievable throughput might change with the location of the interferers, but the trends observed in this section will remain similar.  Next, Fig. \ref{fig:cappos} demonstrates the effect of the RIS position on the throughput. In this simulation, ${\rm Tx}_0$ position is $(2,0)$ and the RIS position is $(d_\alpha,0)$. As we can see, it performs better when it is either close to the ${\rm Rx}_0$ or the ${\rm Tx}_0$. Finally, Fig. \ref{fig:capfwd} shows that if Assumption \ref{assumption:scatter} holds, some throughput enhancement can be achieved with RIS optimization when the carrier frequency is selected from the bands with absorption peaks. Higher absorption coefficients imply more Rayleigh-like channels as the Rician factor decreases. The receive beamformer and RIS can utilize this randomness to extract better performance. Both Fig. \ref{fig:cappos} and \ref{fig:capfwd} are generated with $50$ RIS elements.

\section{Conclusions}
In this paper, we investigated the sensitivity of an RIS-assisted uplink network with respect to both the manifestations of molecular re-radiation in a THz channel. For this setup, the main challenge lies in the joint optimization of the RIS configuration vector and receive beamforming weight under both the manifestations of the re-radiation in the THz channel. To handle that, we developed a parametric THz channel model and proposed an alternating optimization method based on the BCD algorithm. The proposed algorithm breaks the original problem into two smaller sub-problems, and solves them iteratively to converge to an efficient solution. In addition, we also analyzed the convergence of the proposed algorithm. Finally, we numerically compare the performance of this algorithm with a random RIS configuration for different assumptions about the nature of the molecular re-radiation and the existence of a direct link path. To the best of our knowledge, this is the first work that investigates the gap in performance of an RIS-assisted network with optimized configuration, arising from the different assumptions about the molecular re-radiation. Further extensions of the work including the deployment of an RIS with active elements and the presence of errors in CSI estimation are left as future work. 
\appendix
\subsection{Proof of Lemma \ref{lem:NoiseRISdependence}} \label{sec:Lem1Proof}
From (\ref{eq:signalmodel0}), the signal power for the ${\rm Tx}_i$ along the direct path is $P_i\left(\frac{c}{4\pi f d_i}\right)^2$ and subsequently the molecular absorption noise variance due to the direct path would be $\sigma_{m_1,i}^2=\left(\frac{c}{4\pi f d_i}\right)^2P_i[1-\tau(f,d_i)]$.

If the distances between ${\rm Tx}_i$ to RIS, and RIS to ${\rm Rx}_0$ are $d_{\gamma_i}$ and $d_\alpha$ respectively, we inspect the signal $x$ of power $P_i$ through the $m$-th element of RIS with reflection coefficient $(\alpha_me^{j\theta_m})$ without including path-loss terms for simplicity.

The incident signal on the RIS is $x\sqrt{\tau(f,d_{\gamma_i})}+n_1$ where $n_1\sim\mathcal{CN}(0,P_i(1-\tau(f,d_{\gamma_i})))$ is the additive molecular absorption noise. Ultimately, the reflected signal from RIS is
\begin{align*}
  y=&(x\sqrt{\tau(f,d_{\gamma_i})}+n_1)\alpha_me^{j\theta_m}\sqrt{\tau(f,d_{\alpha})}+n_2.
\end{align*}
As the reflected power from the RIS element is $|\alpha_m|^2P_i$, $n_2\sim\mathcal{CN}(0,|\alpha_m|^2P_i(1-\tau(f,d_{\alpha})))$ is the additive noise for the RIS to ${\rm Rx}_0$ path. So, the noise variance due to both the paths is
\begin{align*}
    &\mathrm{E}[|n_1\alpha_me^{j\theta_m}\sqrt{\tau(f,d_{\alpha})}+n_2|^2]\\
   &=|\alpha_m|^2\tau(f,d_{\alpha})P_i(1-\tau(f,d_{\gamma_i}))+|\alpha_m|^2P_i(1-\tau(f,d_{\alpha})) \\
   &=P_i|\alpha_m|^2[1-\tau(f,d_\alpha)\tau(f,d_{\gamma_i})].
\end{align*}

Extending this result to an $N$-element RIS, if the RIS-${\rm Rx}_0$ and ${\rm Tx}_i$-RIS channels are $\mathbf{a}_{h_1}$, and $\mathbf{a}_{h_2}$ with their entries as array factors $a_{h_{1,m}}$ and $a_{h_{2,m}}$ with ULA assumption for RIS, the received signal for SISO is
\begin{align*}
   y
   \!=\!\!x\sqrt{\tau(f,d_{\gamma_i})\tau(f,d_{\alpha})} \!\!\sum\limits_{m=1}^N \!\!\left(\!\alpha_me^{j(\theta_m+a_{h_{1,m}}+a_{h_{2,m}})}\!\right) \!\!+\!\!\! \sum\limits_{m=1}^N\!\! n_m,
\end{align*}
where $n_m\sim\mathcal{CN}(0,P_i|\alpha_m|^2[1-\tau(f,d_\alpha)\tau(f,d_{\gamma_i})])$, and $\sum_{m=1}^Nn_m\sim\mathcal{CN}(0,P_i[1-\tau(f,d_\alpha)\tau(f,d_{\gamma_i})]\sum_{m=1}^N|\alpha_m|^2)$. By including path-loss terms, and writing $\sum_{m=1}^N|\alpha_m|^2$ in matrix form, the molecular noise variance for the reflected signal through RIS can be written as $\sigma_{m_2,i}^2\mathbf{\Theta}_0^H\mathbf{\Theta}_0$ where $\sigma_{m_2,i}^2=\left(\frac{c^2}{16(\pi f)^2 }\frac{1}{d_\alpha d_{\gamma_i}}\right)^2P_i[1-\tau(f,d_\alpha)\tau(f,d_{\gamma_i})]$. The molecular absorption noise variance is then $\zeta\sigma_{m,i}^2$, where $\sigma_{m,i}^2=\sigma_{m_1,i}^2+\sigma_{m_2,i}^2\mathbf{\Theta}_0^H\mathbf{\Theta}_0$ as this noise will only exist for Assumption \ref{assumption:noise} or $\zeta=1$.



\bibliographystyle{IEEEtran}
\bibliography{hokie}

\begin{thebibliography}{10}
\providecommand{\url}[1]{#1}
\csname url@samestyle\endcsname
\providecommand{\newblock}{\relax}
\providecommand{\bibinfo}[2]{#2}
\providecommand{\BIBentrySTDinterwordspacing}{\spaceskip=0pt\relax}
\providecommand{\BIBentryALTinterwordstretchfactor}{4}
\providecommand{\BIBentryALTinterwordspacing}{\spaceskip=\fontdimen2\font plus
\BIBentryALTinterwordstretchfactor\fontdimen3\font minus
  \fontdimen4\font\relax}
\providecommand{\BIBforeignlanguage}[2]{{%
\expandafter\ifx\csname l@#1\endcsname\relax
\typeout{** WARNING: IEEEtran.bst: No hyphenation pattern has been}%
\typeout{** loaded for the language `#1'. Using the pattern for}%
\typeout{** the default language instead.}%
\else
\language=\csname l@#1\endcsname
\fi
#2}}
\providecommand{\BIBdecl}{\relax}
\BIBdecl

\bibitem{tripathi2021millimeter}
S.~Tripathi, N.~V. Sabu, A.~K. Gupta, and H.~S. Dhillon, ``{Millimeter-wave and
  Terahertz Spectrum for 6G Wireless},'' in \emph{6G Mobile Wireless Networks},
  Y.~Wu, S.~Singh, T.~Taleb, A.~Roy, H.~S. Dhillon, M.~R. Kanagarathinam, and
  A.~De, Eds.\hskip 1em plus 0.5em minus 0.4em\relax Springer, 2021.

\bibitem{tataria20216g}
H.~Tataria, M.~Shafi, A.~F. Molisch, M.~Dohler, H.~Sj{\"o}land, and
  F.~Tufvesson, ``6{G} {W}ireless {S}ystems: Vision, {R}equirements,
  {C}hallenges, {I}nsights, and {O}pportunities,'' \emph{Proc., of the IEEE},
  2021.

\bibitem{molisch1998radiation}
A.~F. Molisch and B.~P. Oehry, \emph{{Radiation Trapping in Atomic
  Vapours}}.\hskip 1em plus 0.5em minus 0.4em\relax Oxford University Press,
  1998.

\bibitem{discussion}
J.~Kokkoniemi, J.~Lehtom{\"a}ki, and M.~Juntti, ``{A Discussion on Molecular
  Absorption Noise in the Terahertz Band},'' \emph{Nano commun. networks},
  vol.~8, pp. 35--45, June 2016.

\bibitem{jornet_2012}
J.~M. Jornet and I.~F. Akyildiz, \emph{{Fundamentals of Electromagnetic
  Nanonetworks in the Terahertz Band}}.\hskip 1em plus 0.5em minus 0.4em\relax
  Now Foundations and Trends, 2013.

\bibitem{Harde1}
H.~Harde and D.~Grischkowsky, ``{Coherent Transients Excited by Subpicosecond
  Pulses of Terahertz Radiation},'' \emph{J. Opt. Soc. Am. B}, vol.~8, no.~8,
  pp. 1642--1651, Aug. 1991.

\bibitem{Harde2}
H.~Harde, R.~Cheville, and D.~Grischkowsky, ``{Terahertz Studies of
  Collision-Broadened Rotational Lines},'' \emph{J. Phys. Chem. A}, vol. 101,
  no.~20, pp. 3646--3660, May 1997.

\bibitem{molabs}
S.~A. Hoseini, M.~Ding, M.~Hassan, and Y.~Chen, ``{Analyzing the Impact of
  Molecular Re-Radiation on the MIMO Capacity in High-Frequency Bands},''
  \emph{IEEE Trans. on Veh. Technology}, vol.~69, no.~12, pp. 15\,458--15\,471,
  Dec. 2020.

\bibitem{wu2019beamforming}
Q.~Wu and R.~Zhang, ``{Beamforming Optimization for Wireless Network Aided by
  Intelligent Reflecting Surface with Discrete Phase Shifts},'' \emph{IEEE
  Trans. on Commun.}, vol.~68, no.~3, pp. 1838--1851, Dec. 2019.

\bibitem{RISCR}
J.~Yuan, Y.-C. Liang, J.~Joung, G.~Feng, and E.~G. Larsson, ``{Intelligent
  Reflecting Surface-Assisted Cognitive Radio System},'' \emph{IEEE Trans. on
  Commun.}, vol.~69, no.~1, pp. 675--687, Oct. 2020.

\bibitem{ye2020joint}
J.~Ye, S.~Guo, and M.-S. Alouini, ``{Joint Reflecting and Precoding Designs for
  SER Minimization in Reconfigurable Intelligent Surfaces Assisted MIMO
  Systems},'' \emph{IEEE Trans. on Wireless Commun.}, vol.~19, no.~8, pp.
  5561--5574, May 2020.

\bibitem{VRTHz1}
Y.~Pan, K.~Wang, C.~Pan, H.~Zhu, and J.~Wang, ``{Sum Rate Maximization for
  Intelligent Reflecting Surface Assisted Terahertz Communications},''
  \emph{arXiv preprint arXiv:2008.12246}, 2020.

\bibitem{VRTHz2}
X.~Ma, Z.~Chen, W.~Chen, Y.~Chi, Z.~Li, C.~Han, and Q.~Wen, ``{Intelligent
  Reflecting Surface Enhanced Indoor Terahertz Communication Systems},''
  \emph{Nano Commun. Networks}, vol.~24, p. 100284, Feb. 2020.

\bibitem{VRTHz3}
C.~Chaccour, M.~N. Soorki, W.~Saad, M.~Bennis, and P.~Popovski, ``{Risk-Based
  Optimization of Virtual Reality over Terahertz Reconfigurable Intelligent
  Surfaces},'' in \emph{Proc., IEEE Intl. Conf. on Commun. (ICC)}, July 2020.

\bibitem{SimpleTHz}
J.~Kokkoniemi, J.~Lehtom{\"a}ki, and M.~Juntti, ``{Simple Molecular Absorption
  Loss Model for 200--450 Gigahertz Frequency Band},'' in \emph{European Conf.
  on Networks and Commun. (EuCNC)}, Aug. 2019.

\bibitem{vanExter:89}
M.~van Exter, C.~Fattinger, and D.~Grischkowsky, ``{Terahertz Time-domain
  Spectroscopy of Water Vapor},'' \emph{Opt. Lett.}, vol.~14, no.~20, pp.
  1128--1130, Oct 1989.

\bibitem{Palomar_Eldar_2009}
D.~P. Palomar and Y.~C. Eldar, \emph{{Convex Optimization in Signal Processing
  and Communications}}.\hskip 1em plus 0.5em minus 0.4em\relax Cambridge
  university press, 2010.

\bibitem{cvx}
M.~Grant and S.~Boyd, ``{CVX: Matlab Software for Disciplined Convex
  Programming, version 2.1},'' \url{http://cvxr.com/cvx}, Mar. 2014.

\bibitem{gb08}
------, ``Graph implementations for nonsmooth convex programs,'' in
  \emph{Recent Advances in Learning and Control}, ser. Lecture Notes in Control
  and Information Sciences, V.~Blondel, S.~Boyd, and H.~Kimura, Eds.\hskip 1em
  plus 0.5em minus 0.4em\relax Springer-Verlag Limited, 2008, pp. 95--110.

\bibitem{power2w}
J.~Kokkoniemi, J.~Lehtom{\"a}ki, and M.~Juntti, ``{Stochastic Geometry Analysis
  for Band-Limited Terahertz Band Communications},'' in \emph{Proc., IEEE Veh.
  Technology Conf. (VTC)}, July 2018.

\end{thebibliography}

\end{document}